\documentclass[preprint, superscriptaddress,preprintnumbers, amsmath, epsfig, floatfix, prl]{revtex4}

\usepackage{graphicx}
\usepackage{dcolumn}
\usepackage{bm}
\usepackage{ulem}

\usepackage{times}

\usepackage[utf8]{inputenc}
\usepackage[english]{babel}
\usepackage{keyval}
\graphicspath{{figs/}{figs:}{\figs}}
\setkeys{Gin}{width=0.90\columnwidth}

\usepackage[usenames,dvipsnames]{color}
\usepackage[]{ulem}

\newcommand{\comment}[1]{\textcolor{red}{#1}}
\renewcommand{\comment}[1]{\relax}

\newcommand{\todelete}[1]{\textcolor{green}{\sout{#1}}}
\renewcommand{\todelete}[1]{\relax}

\begin{document}
	
	\title{Emergent room-temperature ferroelectricity in spark-plasma sintered DyCrO$_3$ and LaCrO$_3$}
	\date{\today}
	\author{Suryakanta Mishra}
	\affiliation{Department of Physics, Indian Institute of Technology Kharagpur, Kharagpur-721302, India}
	\author{Keerthana}
	\affiliation{Cryogenic Engineering Centre, Indian Institute of Technology Kharagpur, Kharagpur-721302, India}
	\author{Krishna Rudrapal}
	\affiliation{Advanced Technology Development Centre, Indian Institute of Technology Kharagpur, Kharagpur-721302, India}
	\author{Biswajit Jana}
	\affiliation{Materials Science Centre, Indian Institute of Technology Kharagpur, Kharagpur-721302, India}
	\author{Kazi Parvez Islam}
	\affiliation{Department of Physics, Indian Institute of Technology Kharagpur, Kharagpur-721302, India}
    \author{Archna Sagdeo}
	\affiliation{Accelerator Physics and Synchrotrons Utilization Division, Raja Ramanna Center for Advanced Technology, Indore 452013, India}
	\affiliation{Homi Bhabha National Institute, Training School Complex, Anushakti Nagar, Mumbai 400094, India}
    \author{Ayan Roy Chaudhuri}
	\affiliation{Advanced Technology Development Centre, Indian Institute of Technology Kharagpur, Kharagpur-721302, India}
	\affiliation{Materials Science Centre, Indian Institute of Technology Kharagpur, Kharagpur-721302, India}
    \author{Venimadhav Adyam}
    \affiliation{Cryogenic Engineering Centre, Indian Institute of Technology Kharagpur, Kharagpur-721302, India}
	\author{Debraj Choudhury}
	\email{debraj@phy.iitkgp.ac.in}
	\affiliation{Department of Physics, Indian Institute of Technology Kharagpur, Kharagpur-721302, India}
	
	\begin{abstract}

Identification of novel multiferroic materials with high-ordering temperatures remains at the forefront of condensed matter physics research. In this regard, the antiferromagnetic $\it{R}$CrO$_3$ compounds (like GdCrO$_3$) constitute a promising class of multiferroic compounds, which, however, mostly become ferroelectric concomitant with the antiferromagnetic ordering much below room-temperature, arising from a subtle competition between the ferroelectric off-centering mode and a non-polar antiferrodistortive rotation mode that inhibits ferroelectricity. Recently, room-temperature ferroelectricity of structural origin, arising from off-centering displacements of R and Cr ions, has been identified in spark-plasma sintered GdCrO$_3$ [Suryakanta Mishra et al., Phys. Rev. B \textbf{104}, L180101 (2021)]. Interestingly, some of the experimentally observed non-ferroelectric \textit{R}CrO$_3$ compounds have been theoretically predicted to host similar ferroelectric instabilities. Here, we have identified two such non-ferroelectric \textit{R}CrO$_3$ compounds, one DyCrO$_3$ (which is reported as a quantum paraelectric) and another LaCrO$_3$ (which is paraelectric), and using a modified synthesis protocol involving spark-plasma-sintering (SPS), we have been successful in engineering an intrinsic room-temperature ferroelectricity in the paramagnetic state, driven by non-centrosymmetric structural phase in both SPS-sintered DyCrO$_3$ and LaCrO$_3$, in contrast to room-temperature paraelectricity in solid-state synthesized DyCrO$_3$ and LaCrO$_3$. While the ferroelectricity in SPS-prepared DyCrO$_3$ and LaCrO$_3$ is stable at room-temperature, it undergoes an irreversible transition from a ferroelectric (\textit{P}na2$_1$) phase to a paraelectric (\textit{P}bnm) phase at  $\approx$ 440 K. Significantly, SPS-sintered LaCrO$_3$, which undergoes antiferromagnetic ordering at  $\approx$ 290 K, emerges as a promising near room-temperature multiferroic material.

	\end{abstract}

	\maketitle
	\section{I. Introduction}

	
	\begin{figure}[b!]
		\vspace*{-0.0 in}
		\hspace*{-0.12 in}\scalebox{0.5}
		{\includegraphics{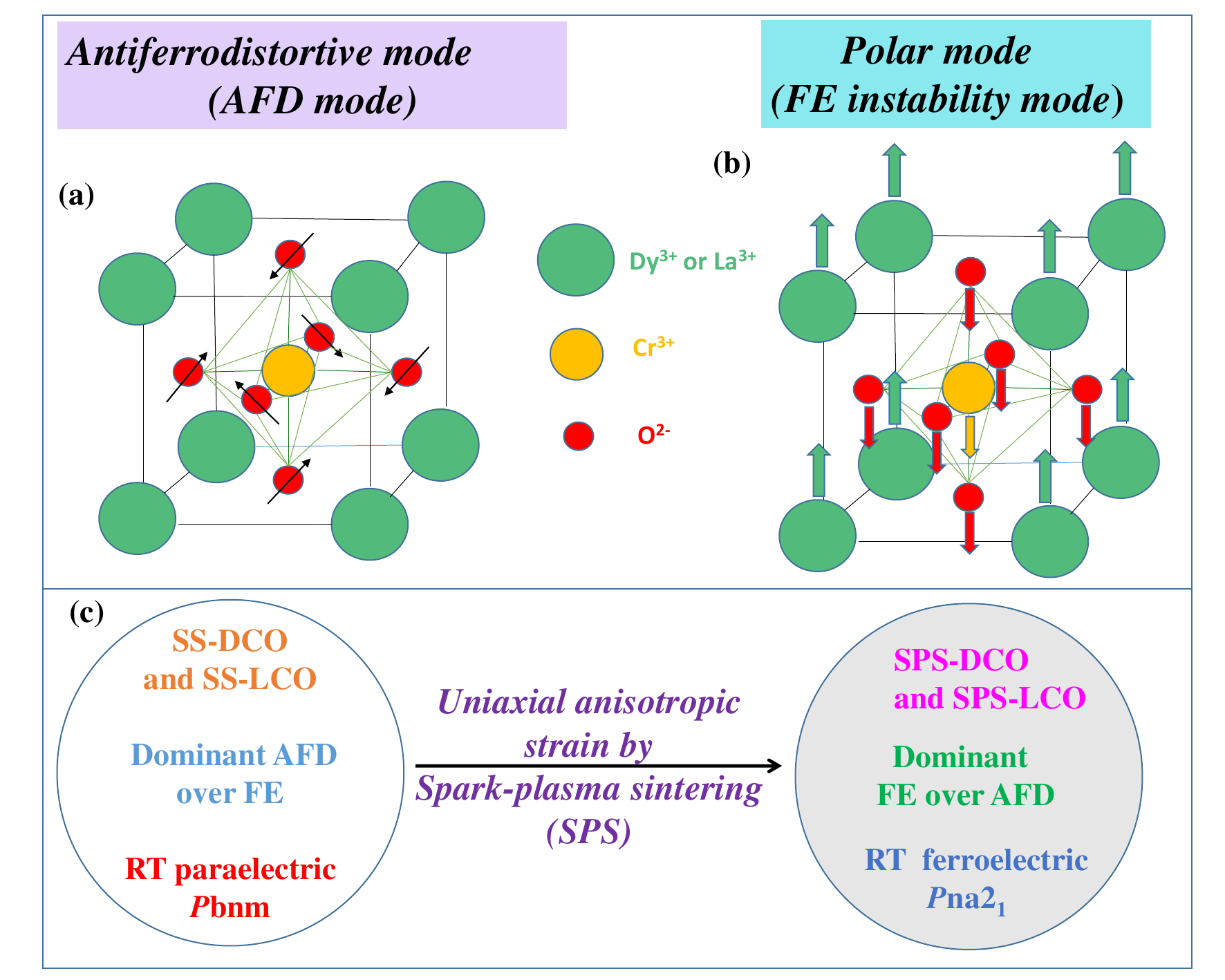}}
		\vspace*{-0.05 in}\caption{(Color online) Schematic visualization of (a) antiferrodistortive mode (AFD), and (b) polar mode (FE instability mode). (c) With the application of uniaxial anisotropic strain on SS-DCO and SS-LCO during spark plasma sintering (SPS), the FE mode seems to get stabilized compared to AFD mode, resulting in room-temperature ferroelectricity arising from noncentrosymmetric \textit{P}na2$_1$ phase in SPS-DCO and SPS-LCO. The following abbreviations stand for: DCO-DyCrO$_3$, LCO-LaCrO$_3$, SS-Solid state synthesized.}\label{Schematics}
	\end{figure}

Multiferroic materials, which are ferroelectric (FE) and spontaneously magnetic (ferromagnetism or antiferromagnetism), are promising for fundamental and applied condensed matter physics research \cite{HSchmid1994, DKhomskii2009, YTokura2003, JFScott2006, JFScott2007, RRamesh2010, MMVopson2015}. Large magnetoelectric coupling in multiferroic materials necessitates the simultaneous emergence of FE and spontaneous magnetic ordering from the same structural unit, which, however, are usually contraindicated in most materials. For example, the "$\it{d}^0$-ness", i.e. 3$\it{d}^0$ (and hence diamagnetic) character of the transition metal $\it{B}$ cation in $\it{AB}$O$_3$ perovskite-related compounds has often been stressed to be necessary to give rise to FE arising from co-operative off-center displacement of the $\it{B}$-cations away from the negative charge center within the corresponding $\it{B}$O$_6$ octahedral cages. FE occurs, for instance, in BaTiO$_3$ because the Ti$^{4+}$(3$\it{d}^0$) ions in the corresponding TiO$_6$ octahedra are cooperatively off-centered \cite{HKrakauer1990, RECohen1992, KMRabe1999}. In case of $\it{AB}$O$_3$ compounds containing non-$\it{d}^0$ $\it{B}$ cation, competing non-polar lattice instabilities related to antiferrodistortive (AFD) rotation modes of the $\it{B}$O$_6$ octahedra become energetically more favourable and thus compete with the polar off-centering mode (schematic visualizations of the AFD and FE instability modes are shown in Figs.\ref{Schematics} (a) and (b)) \cite{PGhosez2009, DVanderbilt1995, KMRabe2009, HYHwang2020, LQChen2015, SMSelbach2016, DVanderbilt2013, RKovacik2011}. Following theoretical predictions \cite{PGhosez2009, NASpaldin2009, KMRabe2010}, it has now been verified experimentally that through appropriate lattice strain it becomes possible to tilt the energy balance away from the non-polar AFD modes and favour the polar off-centering mode in some ferroelectric $\it{A}$MnO$_3$ ($\it{A}$=Sr, Ba, Ca) compounds containing non-$\it{d}^0$ Mn$^{4+}($3$\it{d}^3$) ion \cite{YTokura2011, SLee2021, MFiebig2012}. Large Born-effective charges and off-centering were detected for the Mn$^{4+}$ ion in these $\it{A}$MnO$_3$ compounds, suggesting their dominant contribution to the observed novel FE \cite{PGhosez2009, NASpaldin2009}.

The competition between AFD and FE instability modes is also very common in orthorhombic $\it{R}$CrO$_3$ \cite{CNRRao2005, DTopwal2017}, which constitute an emerging class of multiferroic compounds. For example, FE is observed to arise concomitant with the antiferromagnetic (AFM) ordering at $\approx$170 K in standard solid-state sintered (SS) GdCrO$_3$ \cite{BRajeswaran2012}. Similar ordering temperature for FE and AFM in most $\it{R}$CrO$_3$ (although much below room temperature) has led to contrasting reports in regards to the origin of FE, i.e. whether of structural or magnetic origin \cite{BRajeswaran2012}. Recently, by applying uniaxial pressure at high-temperatures through spark plasma sintering (SPS), we have stabilized FE at room-temperature in SPS GdCrO$_3$, which still undergoes AFM ordering below $\approx$170 K \cite{DChoudhury2021}. FE in SPS GdCrO$_3$, thus, clearly has a structural origin in the non-centrosymmetric \textit{P}na2$_1$ space group (also responsible for FE in SS GdCrO$_3$ below $\approx$170 K) that involves polar, though opposite, off-center displacements for Cr and Gd ions \cite{DTopwal2017, DChoudhury2021}. Some members of $\it{R}$CrO$_3$ family are, however, non ferroelectrics experimentally, such as DyCrO$_3$, which is a quantum-paraelectric \cite{YPSun2018} (quantum fluctuations and AFD instabilities suppress FE order at low temperatures \cite{HBurkard1979}), and LaCrO$_3$, which is a paraelectric \cite{CNRRao2007, YIshii2017, ZGYe2007}. Interestingly, although first-principles calculations deduce large Born-effective charges for La and Cr ions in LaCrO$_3$ (similar to GdCrO$_3$ and some other RCrO$_3$ compounds) that suggest an incipient FE instability \cite{DTopwal2017, RKovacik2011, UVWaghmare2008, DVanderbilt2012}, experimentally FE state has never been realized in DyCrO$_3$ or in LaCrO$_3$ at any temperatures. Here, we show that by adopting a dual synthesis protocol involving SPS, room-temperature FE can be engineered in DyCrO$_3$ (also in LaCrO$_3$) much above the corresponding AFM ordering temperature. Room-temperature intrinsic FE is verified using PUND (positive up–negative down) and PFM (piezoresponse force microscopy) measurements. The net FE distortion at room temperature due to stabilization of the noncentrosymmetric \textit{P}na2$_1$ phase involves dominant off-center displacements of $\it{R}$$^{3+}$ ions in opposite direction to the Cr$^{3+}$ off-center displacements. The estimated ferroelectric polarization, obtained using the atomic positions deduced from Rietveld refinement of corresponding synchrotron x-ray diffraction (XRD) data, is found to be in excellent agreement with the FE polarization values obtained from PUND experiments. Once synthesized, the obtained room-temperature FE \textit{P}na2$_1$ structure is stable, except against further heating under ambient pressures beyond $\approx$450 K, where it converts irreversibly to the centrosymmetric $\it{P}$bnm phase (verified using dielectric, calorimetric and synchrotron XRD investigations). Solid-state synthesized (SS) DyCrO$_3$ and LaCrO$_3$ are found to be paraelectrics at room-temperature, which is in line with the previous literature. SPS synthesized LaCrO$_3$, which is FE at room-temperature, undergoes magnetic ordering at 290 K, thus, coming very close to becoming the first room-temperature multiferroic material in this promising class of $\it{R}$CrO$_3$ compounds.

	\section{II. Experimental Methodology}

Standard solid-state synthesis was used to prepare SS DyCrO$_3$ (SS-DCO) and SS LaCrO$_3$ (SS-LCO) \cite{DChoudhury2021, DTopwal2018}. For this, stoichiometric mixtures of Dy$_2$O$_3$, La$_2$O$_3$ and Cr$_2$O$_3$ were well ground and then calcined in two steps in air, first at 1300$^{\circ}$C and then at 1400$^{\circ}$C for 24 hours. Some part of SS-DCO and SS-LCO, thus prepared, were subjected to spark-plasma sintering (SPS) at 1300$^{\circ}$C for 15 minutes under 60 MPa pressure to obtain SPS-DCO and SPS-LCO. Room-temperature synchrotron powder X-ray diffraction (XRD) using a monochromatic X-ray beam of $\lambda$ = 0.723 {\AA} for DCO (SS-DCO, SPS-DCO, and SPS-DCO-ANN) and 0.721 {\AA} for LCO (SS-LCO, SPS-LCO, and SPS-LCO-ANN) were carried out for structural phase characterizations. Rietveld refinements of room-temperature synchrotron XRD powder diffraction data were performed using FullProf software. Micro Raman measurements were performed using a 514 nm laser at room-temperature. The ferroelectric P-E loop and PUND (positive up–negative down) measurements were conducted using a Radiant P-E loop tracer \cite{JTEvans2011, DChoudhury2020}. Piezoresponce force microscopy (PFM) measurements were also performed with 10 V, 150 Hz ac bias in contact mode to visualize polar domains. The dc magnetic and dielectric permittivity measurements were carried out with the help of a superconducting quantum interference device(SQUID) magnetometer and a LCR meter, respectively. Differential Scanning Calorimetric (DSC) measurements were also performed both in heating and cooling cycles to investigate the phase transition temperature.

\section{III. Results and discussions}


\begin{figure}[b!]
	\vspace*{-0.0 in}
	\hspace*{-0.12 in}\scalebox{0.9}
	{\includegraphics{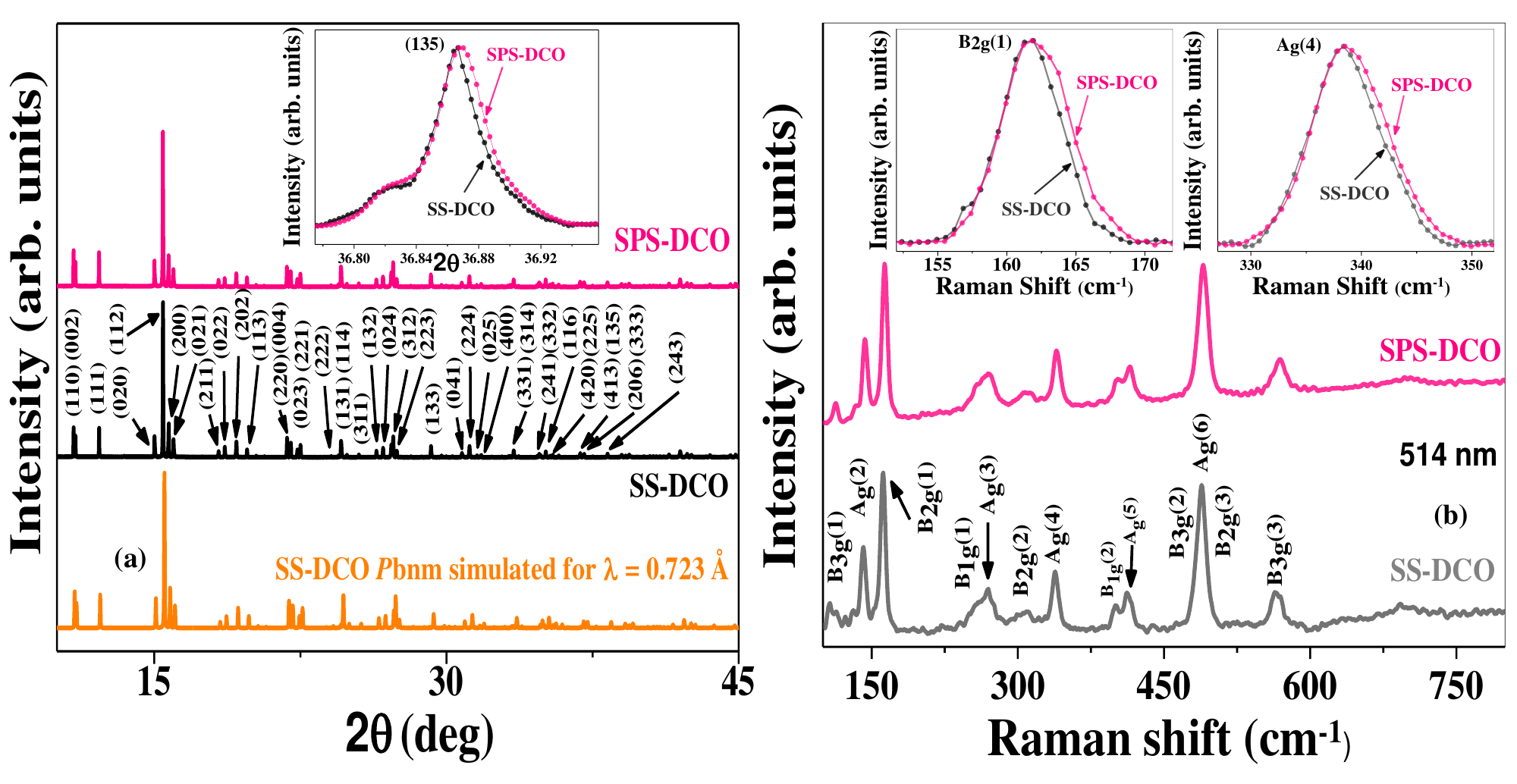}}
	\vspace*{-0.05 in}\caption{(Color online) (a) Room-temperature synchrotron XRD spectra of SPS-DCO and SS-DCO. (b) Room-temperature Raman spectra of SPS-DCO and SS-DCO. The corresponding insets show comparisons of a few of the representative peaks, which suggest that although structurally similar and single-phase, SPS-DCO does exhibit distinct peak-broadenings as compared to SS-DCO.}\label{XRD,Raman}
\end{figure}


\begin{figure}[t]
	\vspace*{-0.0 in}
	\hspace*{-0.12 in}\scalebox{0.6}
	{\includegraphics{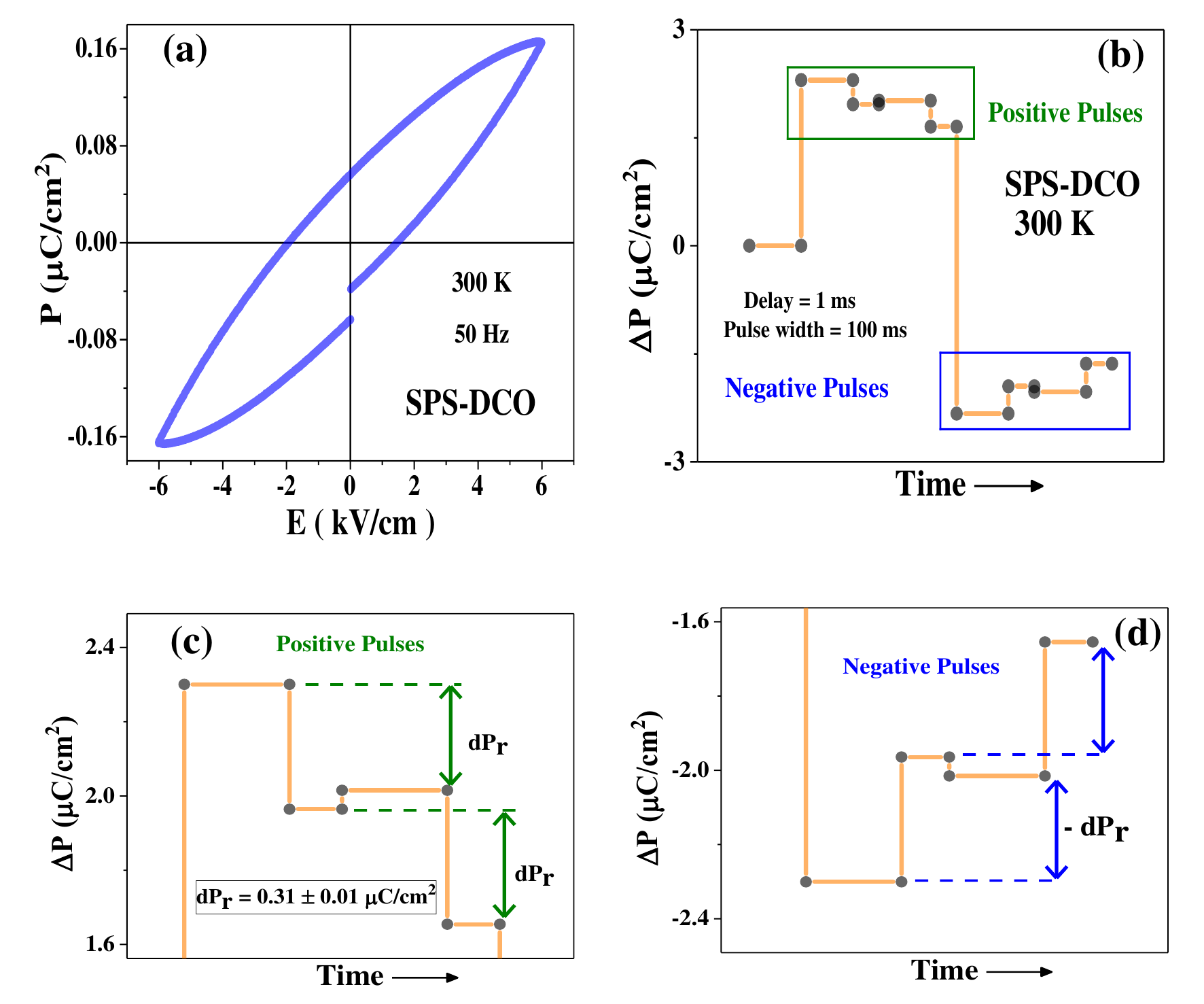}}
	\vspace*{-0.05 in}\caption{(Color online) (a) Room-temperature P-E loop of SPS-DCO collected at a frequency of 50 Hz. (b) Room temperature PUND data at 6 kV/cm applied electric field on SPS-DCO. The figures (c) and (d) show enlarged views of the switchable electric polarization responses under positive and negative electric pulses of PUND data.}\label{PUND}
\end{figure}


\begin{figure}[t]
	\vspace*{-0.0 in}
	\hspace*{-0.12 in}\scalebox{0.8}
	{\includegraphics{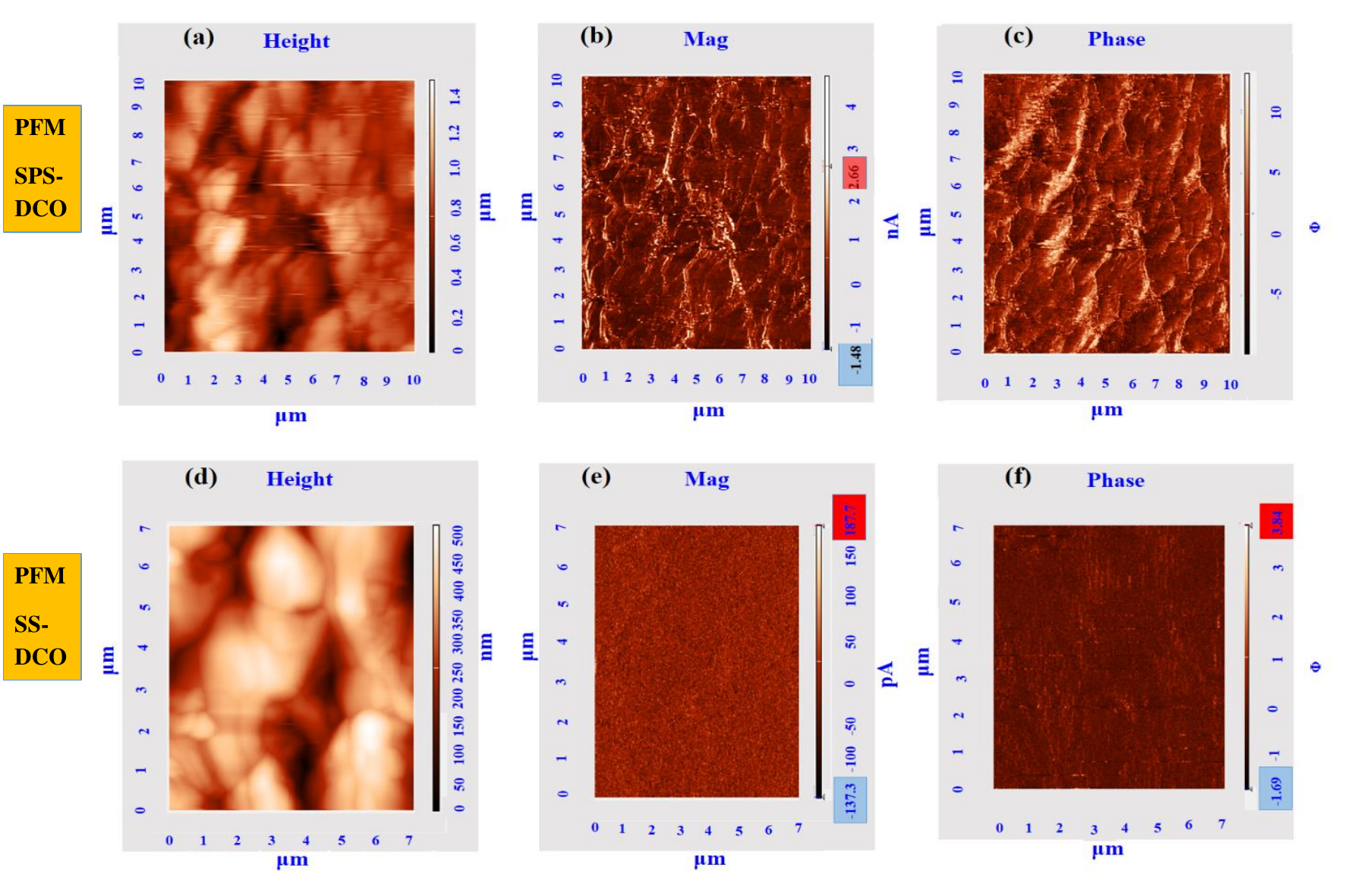}}
	\vspace*{-0.05 in}\caption{(Color online) Room-temperature (a) topographical AFM, (b) amplitude contrast, and (c) phase contrast PFM images (colour contrast present in phase image correspond to the signature of piezoelectricity in the compound) of SPS-DCO. Figures (d)-(f) represent topographical AFM and the absence of amplitude and phase contrast in the PFM images of SS-DCO.}\label{PFM}
\end{figure}


\begin{figure}[h]
	\vspace*{-0.15 in}
	\hspace*{-0.12 in}\scalebox{0.5}
	{\includegraphics{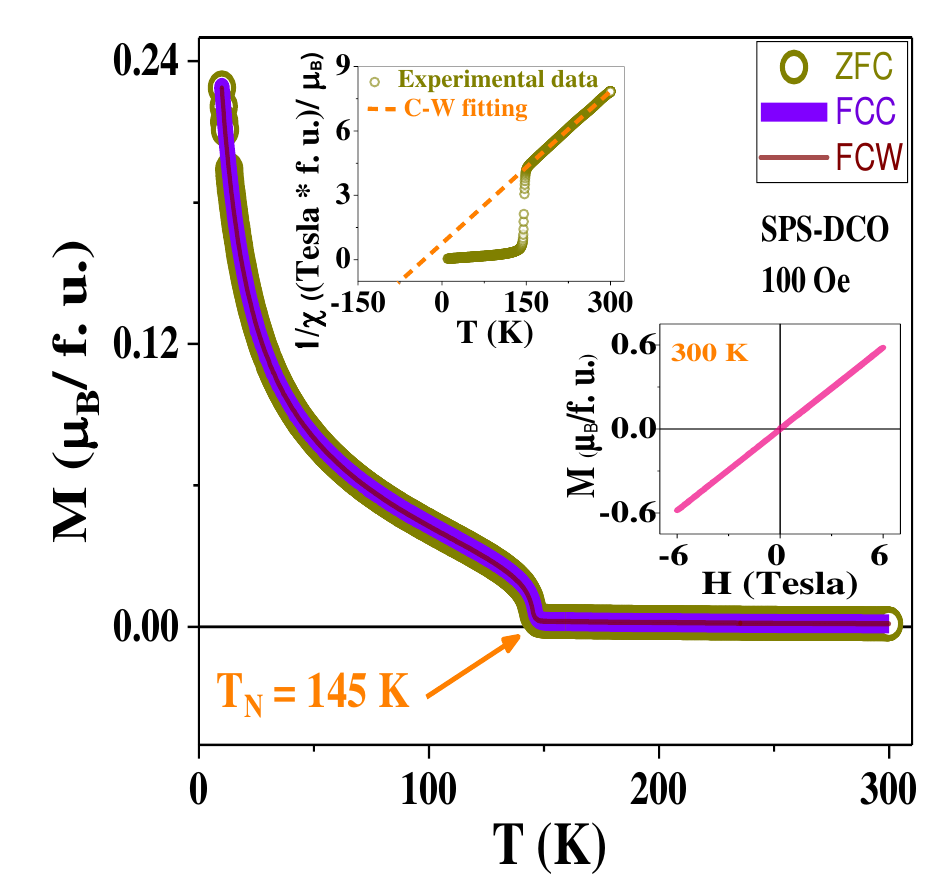}}
	\vspace*{-0.18 in}\caption {(Color online) Temperature ($\it{T}$) dependence of magnetization ($\it{M}$) of SPS-DCO collected in zero-field cooled heating (ZFC), field-cooled cooling (FCC) and field-cooled heating (FCW) modes with applied magnetic field ($\it{H}$) of 100 Oe (the upper inset shows $\it{\chi}^{-1}$-$\it{T}$ data which clearly highlight the antiferromagnetic ordering and, the lower inset illustrates the linear dependence of $\it{M}$ on $\it{H}$ highlighting the paramagnetic state at room temperature).}\label{Magnetic}
\end{figure}

Due to similarities and for the sake of brevity and emphasis, we will largely address the methods and results involving DyCrO$_3$ in this manuscript and will save the discussion of the significant results on LaCrO$_3$ towards the later part of the manuscript. As seen through the synchrotron XRD data in Fig. \ref{XRD,Raman}(a) and room-temperature Raman data in Fig. \ref{XRD,Raman}(b), both SS-DCO and SPS-DCO are single-phase and appear to be structurally similar \cite{EMoran2013, RIWalton2012, MJain2013}. SPS-DCO, however, exhibits distinct peak-broadenings in comparison to SS-DCO both in the XRD and Raman data (mainly in modes related to Dy$^{3+}$ ions and surrounding oxygen ions \cite{RIWalton2012, MJain2013}), origin of which will be discussed in further details later in the manuscript. Interestingly, we find a finite, though somewhat lossy, electric-polarization ($\it{P}$) vs. electric-field ($\it{E}$) loop at room-temperature in SPS-DCO (shown in Fig. \ref{PUND}(a)). In order to verify whether SPS-DCO is FE at room-temperature, we adopted the PUND FE characterization technique since it is a well-established and sensitive tool to extract out intrinsic FE from other extrinsic contributions \cite{BRajeswaran2012, DChoudhury2021, CNRRAO2014, KHKim2012, KHKim2010, DChoudhury2020}. Remarkably, the room-temperature PUND results, as shown in Figs. \ref{PUND}(b)-(d), confirm the existence of finite, switchable intrinsic FE remanent polarization (dP$_{\rm{r}}$= 0.31 $\pm$ 0.01 $\mu$C/cm$^{2}$) in SPS-DCO. Also, as seen in Figs. \ref{PFM}(a)-(c), different amplitude and phase contrast regions (corresponding to different ferroelectric domains) are identifiable in the piezoresponse-force microscopy (PFM) data on SPS-DCO at room temperature. While, due to the lossy nature of SS-DCO, reliable PUND measurements could not be carried out, PFM data on SS-DCO clearly show the absence of any piezoresponse, as seen in  Figs. \ref{PFM}(d)-(f), elucidating the room-temperature paraelectric state in SS-DCO in contrast to the FE SPS-DCO.


   \begin{figure}[b!]
   	\vspace*{-0.0 in}
   	\hspace*{-0.0 in}\scalebox{0.6}
   	{\includegraphics{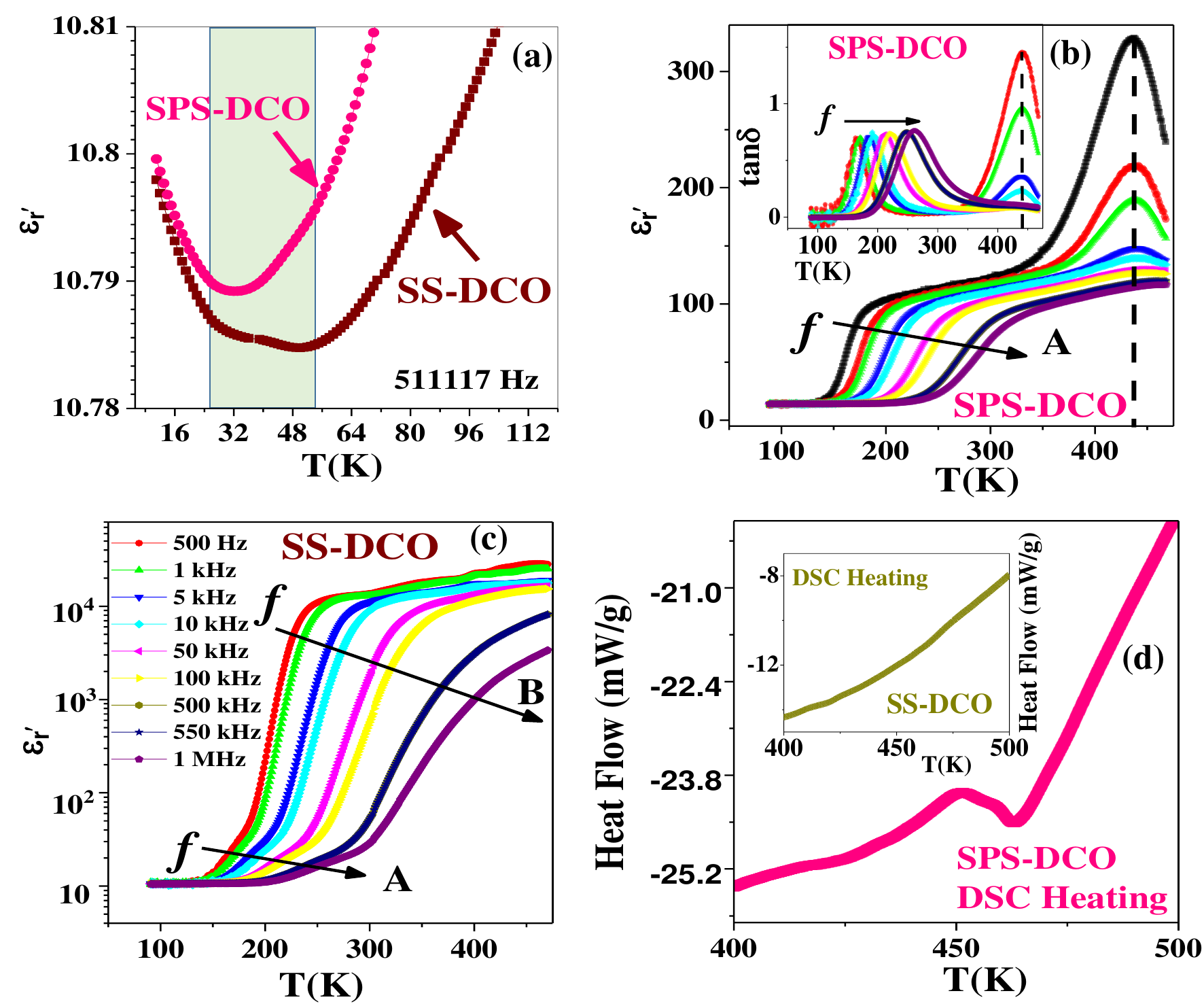}}
   	\vspace*{-0.0 in}\caption{(Color online) (a) Comparison of relative dielectric permittivity (${\epsilon}^{'}_{\rm{r}}$) of SPS-DCO and SS-DCO collected at 511117 Hz. Temperature ($\it{T}$) dependencies of relative dielectric permittivity (${\epsilon}^{'}_{\rm{r}}$) at various electric field frequencies of (b) SPS-DCO and (c) SS-DCO. The inset to (b) shows corresponding dielectric loss data of SPS-DCO. (d) $\it{T}$ dependent DSC heating data of SPS-DCO highlighting similar higher temperature transition (the inset shows DSC data of SS-DCO, taken in the heating cycle that exhibits absence of any corresponding transition).}\label{Dielectric, DSC}
   \end{figure}


 \begin{figure}[b]
 	\vspace*{-0.15 in}
 	\hspace*{-0.0 in}\scalebox{0.6}
 	{\includegraphics{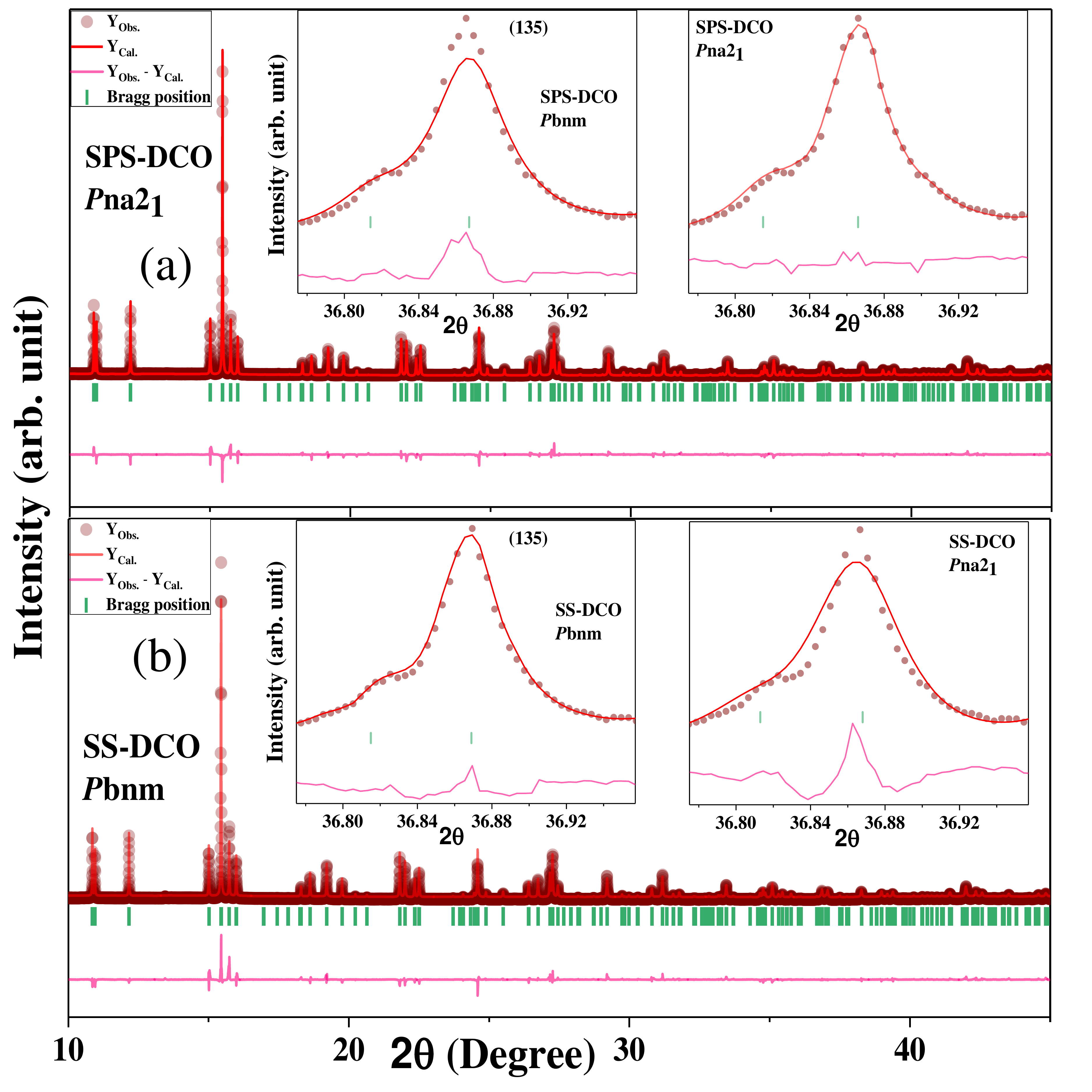}}
 	\vspace*{-0.0 in}\caption{(Color online) (a) The Rietveld refinement of room temperature synchrotron XRD spectra of SPS-DCO considering \textit{P}na2$_1$ space group. The left and right insets show the (135) peak of SPS-DCO, refined with \textit{P}bnm (R$_{\rm{p}}$= 10.4, R$_{\rm{wp}}$= 13.6, $\chi^{2}$= 4.08) and with \textit{P}na2$_1$ (R$_{\rm{p}}$= 9.26, R$_{\rm{wp}}$= 13.0, $\chi^{2}$= 3.73) space groups, respectively. (b) The Rietveld refinement of room temperature synchrotron XRD spectra of SS-DCO with \textit{P}bnm space group. The left and right insets show the (135) peak of SS-DCO, refined with \textit{P}bnm (R$_{\rm{p}}$= 10.5, R$_{\rm{wp}}$= 14.5, $\chi^{2}$= 3.86) and with \textit{P}na2$_1$ (R$_{\rm{p}}$= 11.1, R$_{\rm{wp}}$= 15.3, $\chi^{2}$= 4.30) space groups, respectively.}\label{Refinement}
 \end{figure}


 \begin{figure}[t]
 	\vspace*{-0.15 in}
 	\hspace*{-0.0 in}\scalebox{0.6}
 	{\includegraphics{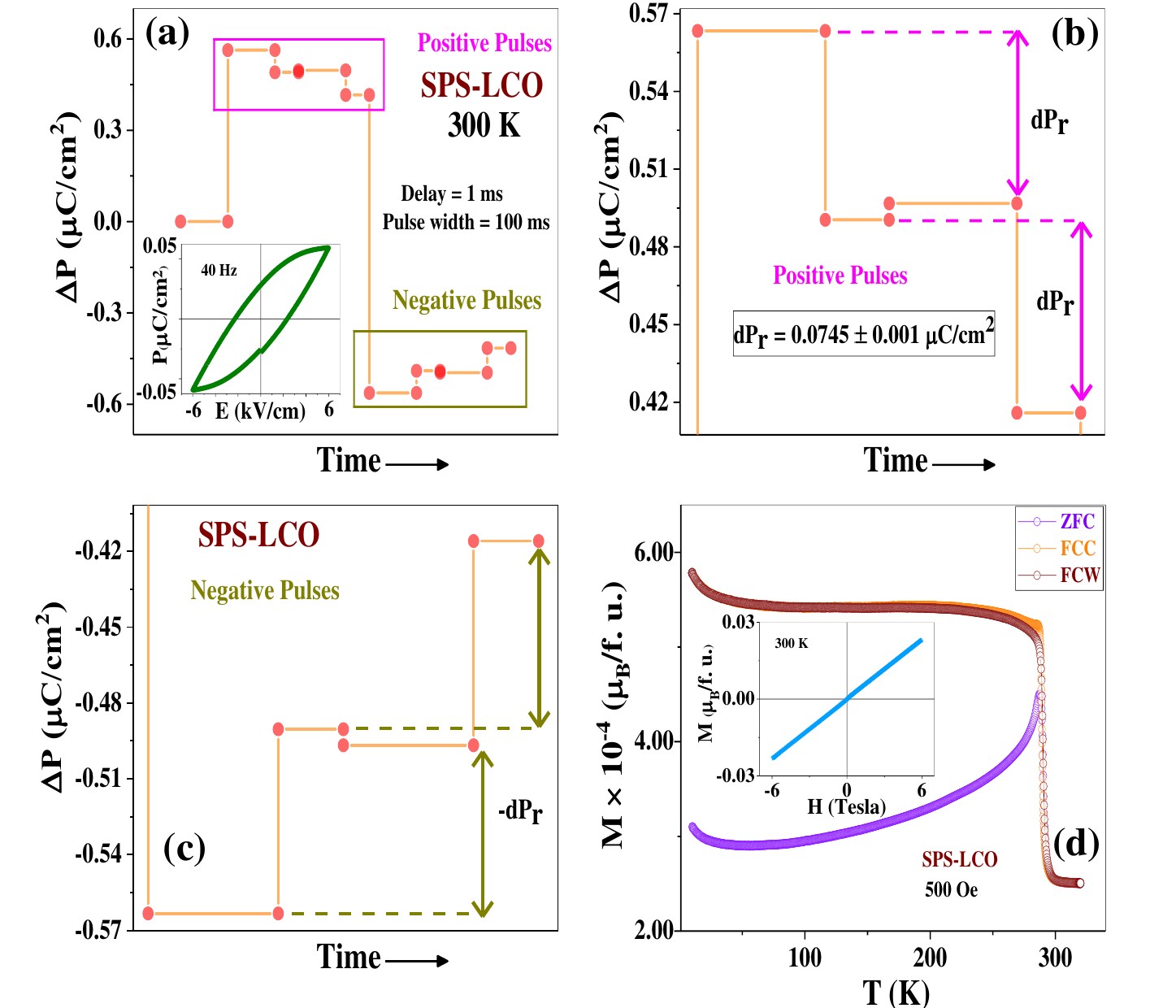}}
 	\vspace*{-0.0 in}\caption{(Color online) (a) Room temperature PUND data of SPS-LCO collected with 6 kV/cm applied electric field (the inset shows the room-temperature P-E loop of SPS-LCO recorded at a frequency of 40 Hz). The figures (b) and (c) show enlarged views of the switchable electric polarization responses under positive and negative electric pulses of PUND data. (d) ZFC, FCC, and FCW M-T data of SPS-LCO indicate an antiferromagnetic transition at around 290 K. Linear M-H data in the corresponding inset indicates a paramagnetic state of SPS-LCO at room temperature}\label{Dielectric, DSC LCO}
 \end{figure}

In order to investigate any role of magnetism to the observed room-temperature FE in SPS-DCO, temperature ($\it{T}$) and magnetic-field ($\it{H}$) dependent magnetization ($\it{M}$) measurements were carried out. As seen through the corresponding $\it{M}$-${T}$ and inverse magnetic-susceptibility $\it{\chi}^{-1}$-$\it{T}$ data of SPS-DCO in Fig. \ref{Magnetic} and its upper inset, SPS-DCO undergoes a paramagnetic (PM) to antiferromagnetic (AFM) transition at $\approx$145 K. This is also consistent with a linear room-temperature $\it{M}$-$\it{H}$ loop (without trace of any hysteresis) of SPS-DCO, as seen in the lower inset of Fig. \ref{Magnetic}. Importantly, all the above magnetic properties of SPS-DCO, including the PM to AFM transition temperature, are near-identical to that of SS-DCO (as seen in Fig. S3 and its insets \cite{SI2022DCO}) \cite{MJain2013, PMohanty2021, MJain2015}. Thus, any role of magnetism to the observed room-temperature FE in SPS-DCO can be clearly ruled out.

Small increment in the relative dielectric permittivity (${\epsilon^{'}}_{\rm{r}}$) values of polycrystalline SS-DCO below $\approx$50 K (illustrated by the shaded region in Fig. \ref{Dielectric, DSC}(a)) seems consistent with a low-temperature increase in ${\epsilon^{'}}_{\rm{r}}$ seen in single-crystalline DyCrO$_3$ (albeit at slightly higher temperatures) and that is reported to arise as a consequence of the quantum paraelectric nature of DyCrO$_3$ \cite{YPSun2018}. For single-crystalline DyCrO$_3$, a small increase ($\approx$0.13) in ${\epsilon^{'}}_{\rm{r}}$ is found to arise only along c-axis below 150 K (along other perpendicular two-axes, ${\epsilon^{'}}_{\rm{r}}$ decreases with lowering of temperature). Understandably, due to averaging effect in polycrystalline samples, the corresponding rise in ${\epsilon^{'}}_{\rm{r}}$ will become weaker and observable only at further lower temperature. Since SS-DCO do not exhibit any other transition around $\approx$50 K, the weak rise in ${\epsilon^{'}}_{\rm{r}}$ below $\approx$50 K is understood to arise due to its reported quantum paraelectric state. At further lower temperatures of $\approx$20 K, spin-reorientation transition of Dy$^{3+}$ spins in DyCrO$_3$ leads to a further increase (seen in Fig. \ref{Dielectric, DSC}(a)) in ${\epsilon^{'}}_{\rm{r}}$ values of DyCrO$_3$ \cite{YPSun2016}. Interestingly, while the rise in ${\epsilon^{'}}_{\rm{r}}$ values is clearly observable below the spin-reorientation transition in case of both SPS-DCO and SS-DCO, any rise in ${\epsilon^{'}}_{\rm{r}}$ values is not observed for SPS-DCO below $\approx$50 K, likely suggesting the melting of the corresponding quantum paraelectric state in SPS-DCO. Further, a clear peak in $\it{T}$-dependent ${\epsilon^{'}}_{\rm{r}}$ and dielectric loss data (collected during heating run under ambient pressure), which does not disperse with varying electric-field frequencies, as seen in Figs. \ref{Dielectric, DSC}(b), clearly suggests that SPS-DCO undergoes a FE to paraelectric (presumably) phase transition at $\approx$440 K. The transition at $\approx$440 K in SPS-DCO also becomes clearly evident in the corresponding DSC data collected on SPS-DCO during the heating run, as seen in Fig. \ref{Dielectric, DSC}(d). Consistent with the room-temperature paraelectric state of SS-DCO, such a high-temperature dispersionless phase transition is not observable in the corresponding ${\epsilon^{'}}_{\rm{r}}$-$\it{T}$ data (which instead exhibits strong Maxwell-Wagner dielectric relaxation, marked as B in Fig. \ref{Dielectric, DSC}(c) \cite{ARvonHippel1996, DChoudhury2012, PNSanthosh2020}; the dielectric relaxation A at lower temperatures, which is similarly found in the case of SPS-DCO is currently being investigated in further details). Similarly, the DSC data of SS-DCO, collected in the heating run, do not exhibit signature of any high-temperature phase transition, as seen in the inset to Fig. \ref{Dielectric, DSC}(d). Importantly, the observed FE in SPS-DCO is found to be reproducible and stable at room-temperature over a gap of many months (the maximum that we have checked for is after a gap of 12 months), as shown in Fig. S4 of \cite{SI2022DCO}.

To investigate whether the obtained room-temperature FE in SPS-DCO is a thermodynamically-stabilized or kinetically-stabilized phase, $\it{T}$-dependent ${\epsilon^{'}}_{\rm{r}}$ in subsequent heating cycles and DSC measurements in subsequent cooling and heating cycles were performed under ambient-pressure condition. Interestingly, the peak in ${\epsilon^{'}}_{\rm{r}}$ in subsequent heating and the corresponding peak in the DSC cooling at $\approx$ 440 K is absent, as seen in Fig. S1 of \cite{SI2022DCO}. To further verify, SPS-DCO had been further annealed at 1300$^{\circ}$ C in air under ambient pressure condition and slowly cooled to room temperature to form SPS-DCO-ANN. Consistent with the paraelectric state of SPS-DCO-ANN at room temperature, corresponding PFM data (shown in Figs. S2(d)-(f)) do not show any phase and amplitude contrast. These measurements, thus, clearly elucidate that room-temperature FE in SPS-DCO is a kinetically arrested phase, that is unstable against heating of the sample beyond $\approx$440 K.

 In order to understand the structural phase responsible for room-temperature FE in SPS-DCO, we refer to our earlier first-principles calculations on the relative energy stability among the various possible structural space groups in the $\it{R}$CrO$_3$ compounds \cite{DChoudhury2021}. Results from our first-principles calculation, which is also in consistence with other similar investigations \cite{DTopwal2017}, suggest that two structural space groups, centrosymmetric \textit{P}bnm and non-centrosymmetric \textit{P}na2$_1$ are energetically more favourable (the energy difference between these two structures is within our calculation error limit) in case of many of the $\it{R}$CrO$_3$ compounds. Consistent with the above, the paraelectric to ferroelectric phase transition in $\it{R}$CrO$_3$ materials have also been ascribed to a phase transition between the $\it{P}$bnm to the $\it{P}$na2$_1$ structures, respectively \cite{SGiri2014, DTopwal2017, SGiri2015}. Accordingly, we have refined the room-temperature synchrotron XRD spectra of SS-DCO and SPS-DCO by adapting both the \textit{P}bnm and the \textit{P}na2$_1$ space groups (refined lattice parameters are shown in Table-I \cite{SI2022DCO}). Interestingly, while the room-temperature XRD spectrum of SPS-DCO can be better fitted using the non-centrosymmetric $\it{P}$na2$_1$ space group, the same for SS-DCO can be better accounted for by adapting the centrosymmetric $\it{P}$bnm space group, as seen in Fig. \ref{Refinement}. In addition, the structure of SPS-DCO-ANN, as determined from room-temperature synchrotron XRD data, is found to be better described with centrosymmetric \textit{P}bnm in consistence with its room-temperature paraelectric state (shown in Figs. S2(a)-(c) of \cite{SI2022DCO}). Using the refined structural parameters of the $\it{P}$na2$_1$ space group for SPS-DCO, the ionic contribution to the ferroelectric polarization was calculated using the formula reported in Refs. \cite{SGiri2014, DChoudhury2021}. Significantly, the calculated ionic contribution to the FE polarization comes out to be $\approx$0.1117 $\pm$ 0.0001 $\mu$C/cm$^2$, which agrees very well with the remanant FE polarization value (i.e. P= 0.155 $\pm$ 0.01 $\mu$C/cm$^2$ where P= $\frac{1}{2}$ dP$_r$) obtained in room-temperature PUND experiments on SPS-DCO.

Further, we also report intrinsic room-temperature ferroelectricity in SPS-LCO (results of room-temperature $\it{P}$-$\it{E}$ and PUND measurements are shown in Figs. \ref{Dielectric, DSC LCO}(a)-(c)), whereas SS-LCO remains paraelectric (supported data are shown in Fig. S6(d) and Fig. S9(b) of \cite{SI2022DCO}), as reported earlier. In consistence with the above, SPS-LCO and SS-LCO are found to crystallize in the noncentrosymmetric \textit{P}na2$_1$ and centrosymmetric \textit{P}bnm phases at room temperature, respectively (corresponding synchrotron XRD and refinement results are shown in Fig. S8, Fig. S9 and Table-II of \cite{SI2022DCO}). Interestingly, both SPS-LCO and also SS-LCO undergo AFM ordering just below room-temperature, at $\approx$290 K ($\it{M}$-$\it{T}$ data of SPS-LCO is shown in Fig. \ref{Dielectric, DSC LCO}(d) and corresponding data of SS-LCO is shown in Fig. S5 of \cite{SI2022DCO}) \cite{CNRRao2007, YIshii2017, ZGYe2007, JBGoodenough2011}. The room-temperature ferroelectricity in SPS-LCO, while reproducible and stable at room-temperature, undergo a similar irreversible FE to paraelectric phase transition (corresponding dispersionless transition in ${\epsilon^{'}}_{\rm{r}}$-$\it{T}$ and DSC data, collected during the heating runs, are included in Figs. S6(a)-(C) of \cite{SI2022DCO}) at high temperatures similar to SPS-DCO (Fig. S7 of \cite{SI2022DCO}). A similarly annealed SPS-LCO, SPS-LCO-ANN (annealed under similar conditions as SPS-DCO-ANN) is found to crystallize in the centrosymmetric \textit{P}bnm phase (corresponding synchrotron XRD data and refinement results are shown in Fig. S10 of  \cite{SI2022DCO}) establishing the kinetic origin of ferroelectricity in SPS-LCO. SPS-LCO, however, exhibits a lower FE polarization as compared to SPS-DCO at room-temperature, likely because the ionic radius of La$^{3+}$ ions is much larger than Dy$^{3+}$, which causes a smaller magnitude of off-center displacement of La$^{3+}$ ion in SPS-LCO as compared to Dy$^{3+}$ ion in SPS-DCO. Also, the calculated ionic contribution to the ferroelectric polarization, using the refined atomic positions (from synchrotron XRD refinement results) of SPS-LCO come out to be $\approx$ 0.01908 $\pm$ 0.0001 $\mu$C/cm$^2$, which is in good agreement with the experimentally determined remanent FE polarization value ($\approx$ 0.0373 $\pm$ 0.001 $\mu$C/cm$^2$) of SPS-LCO. Significantly, SPS-LCO becomes multiferroic below $\approx$290 K, which is the highest in this $\it{R}$CrO$_3$ family and is very close to becoming a room-temperature multiferroic material.

 \section{IV. Summary}
 
In conclusion, we successfully engineer room-temperature ferroelectricity in both DyCrO$_3$ and LaCrO$_3$ when they are synthesized using a modified protocol that involves spark-plasma sintering (SPS) at high temperatures, despite being traditionally classified as quantum-paraelectric and paraelectric materials, respectively. This is the first experimental realization of ferroelectricity (FE) in either of these compounds, despite the theoretical prediction of a ferroelectric (FE) instability over antiferrodistortive (AFD) non-polar modes. In contrast, the solid-state synthesized (SS) DyCrO$_3$ and LaCrO$_3$ are found to remain paraelectrics, following earlier reports. Even though the magnetic properties of the SPS and SS synthesized compounds are the same and they both go through antiferromagnetic ordering at temperatures lower than room temperature, it is important to note that the room-temperature FE in both the SPS synthesized compounds is found to be of structural origin arising from off-centering displacements of rare-earth $\it{R}$ and Cr ions. The kinetically-arrest process during SPS that leads to the room-temperature FE phase likely prefers the FE instability modes over the AFD modes in this emerging class of RCrO$_3$ multiferroic compounds.
	
	\section{V. Acknowledgments}

SM would like to acknowledge the financial support from MoE, India. DC acknowledges STARS, MoE, India (MoE-STARS/STARS-1/238) for financial support. SM and DC would like to acknowledge the use of the Raman spectroscopy and SPS central facilities at IIT Kharagpur. DC would like to acknowledge insightful discussions with Professor D. D. Sarma.

\newpage
\begin{center}
 \section{\underline{SUPPLEMENTARY MATERIAL}}	
\end{center}

\section{VI. Temperature dependencies of relative dielectric permittivity collected in 2nd heating cycle, and DSC data of SPS-DCO in cooling cycle}

 \renewcommand{\thefigure}{S1}
 \begin{figure}[h!]
 	\vspace*{-0.15 in}
 	\hspace*{-0.0 in}\scalebox{0.5}
 	{\includegraphics{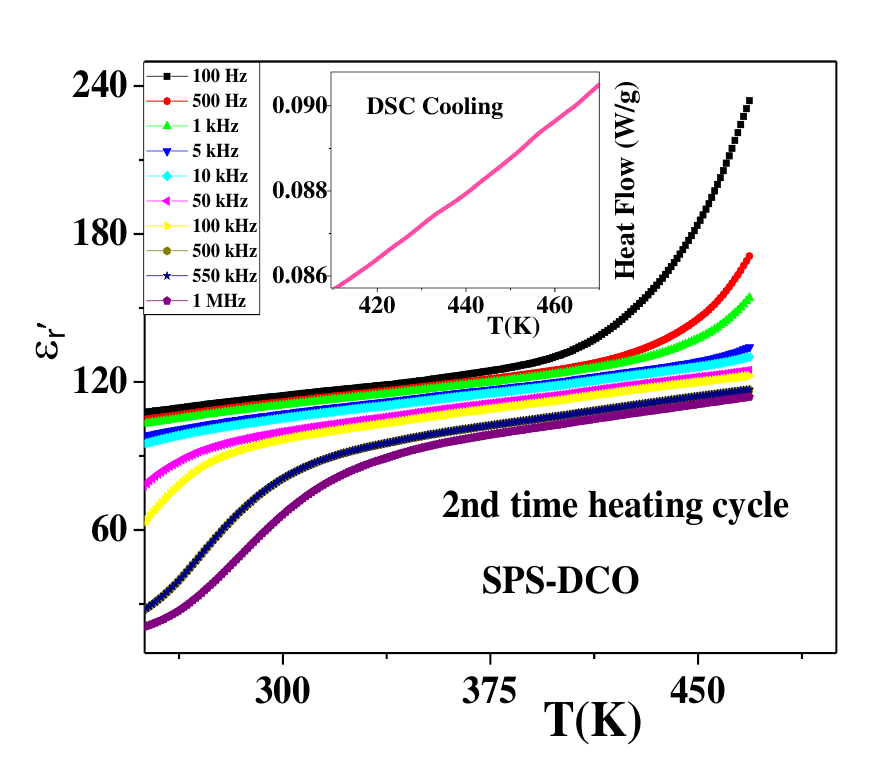}} 
    \vspace*{-0.0 in}\caption{(Color online) The relative dielectric permittivity data of SPS-DCO taken in the 2nd heating cycle which shows the absence of high temperature transition (the inset shows the DSC cooling data of SPS-DCO).}\label{}
 \end{figure}

 \section{VII. Lattice parameters obtained from Rietveld refinement of room temperature synchrotron XRD of SPS- DCO and SS-DCO}

\setlength{\tabcolsep}{0.22 cm}
\begin{table}[h!]
	\centering
	\caption{}.
	\label{}
	\vspace*{0.5cm}
	\begin{tabular}{|c|c|}
			\hline
			           
			SPS-DCO-\textit{P}na2$_1$ &  SS-DCO-\textit{P}bnm  \\
			
			\hline
		                 
			            a = 5.53309(3) {\AA}  &  a = 5.27954(2) {\AA}  \\
			
			\hline
			              
			          b = 5.27900(2) {\AA} & b = 5.53203(3)  {\AA}    \\
			
			\hline
			              
			        c= 7.57270(3) {\AA}  &  c= 7.57316(3) {\AA}\\
			
			\hline
			
	\end{tabular}
	
\end{table}

\newpage

\section{VIII. Room-temperature synchrotron XRD, Rietveld refinement, and room-temperature PFM images of SPS-DCO-ANN sample}

\renewcommand{\thefigure}{S2}
\begin{figure}[h!]
	\vspace*{-0.15 in}
	\hspace*{-0.0 in}\scalebox{0.8}
	{\includegraphics{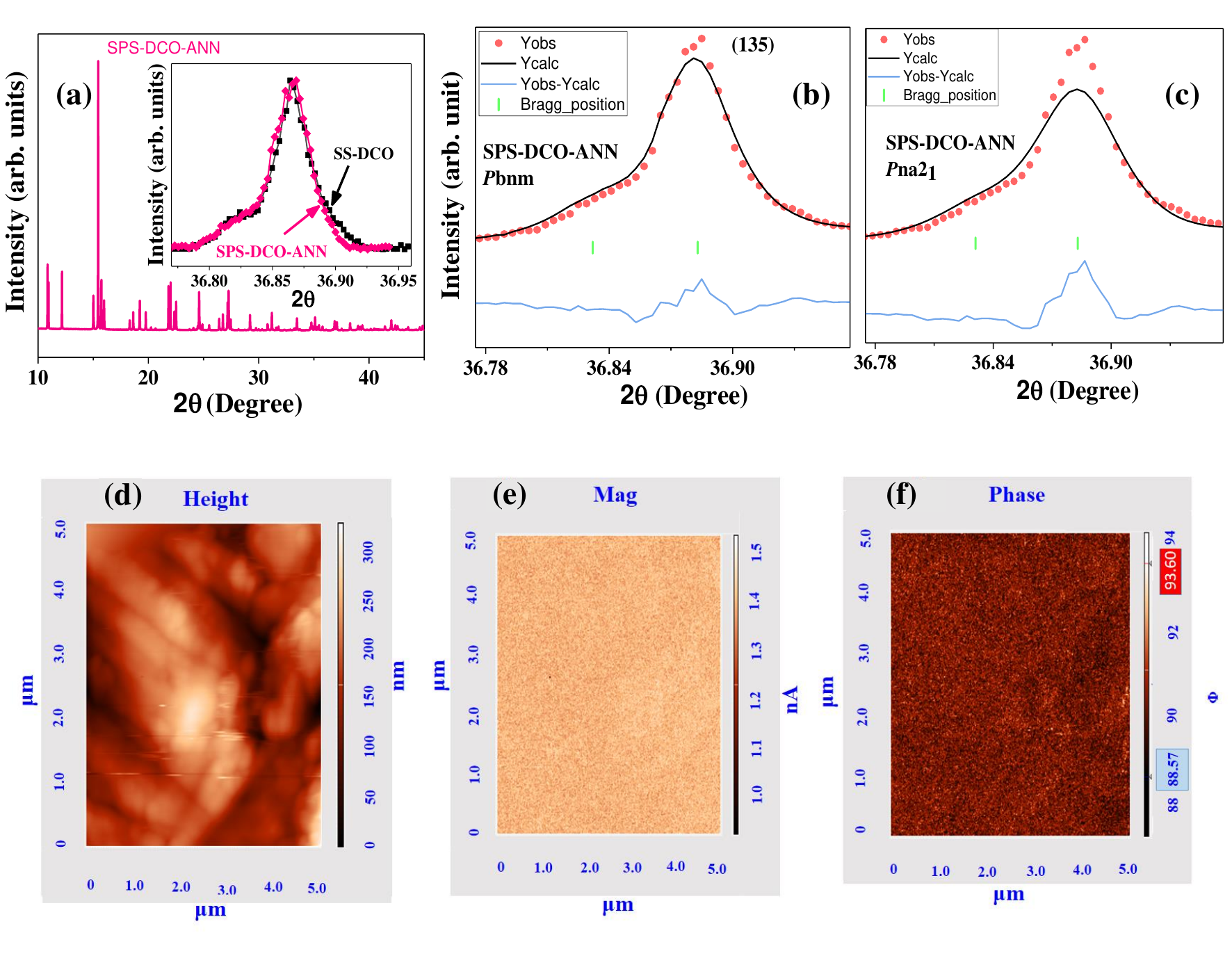}} 
	\vspace*{-0.0 in}\caption{(Color online) (a) Room-temperature synchrotron XRD data of SPS-DCO-ANN sample. The inset shows the XRD peak comparison between SPS-DCO-ANN and SS-DCO. The Rietveld refinement of room temperature synchrotron XRD of SPS-DCO annealed sample (SPS-DCO-ANN) with (b) \textit{P}bnm (R$_{\rm{p}}$ = 9.07, R$_{\rm{wp}}$ = 12.6, $\chi^{2}$ = 3.64), (c) \textit{P}na2$_1$ (R$_{\rm{p}}$ = 9.76, R$_{\rm{wp}}$ = 13.3, $\chi^{2}$  = 4.05). Figures (d)-(f) represent room-temperature topographical AFM, and the absence of any contrast in the amplitude, and phase images of SPS-DCO-ANN}\label{}
\end{figure}

	\newpage
	
	\section{IX. Magnetization (\textit{M}) vs. temperature (\textit{T}) data of SS-DCO}

	\renewcommand{\thefigure}{S3}
	\begin{figure}[h!]
		\vspace*{-0.15 in}
		\hspace*{-0.0 in}\scalebox{0.6}
		{\includegraphics{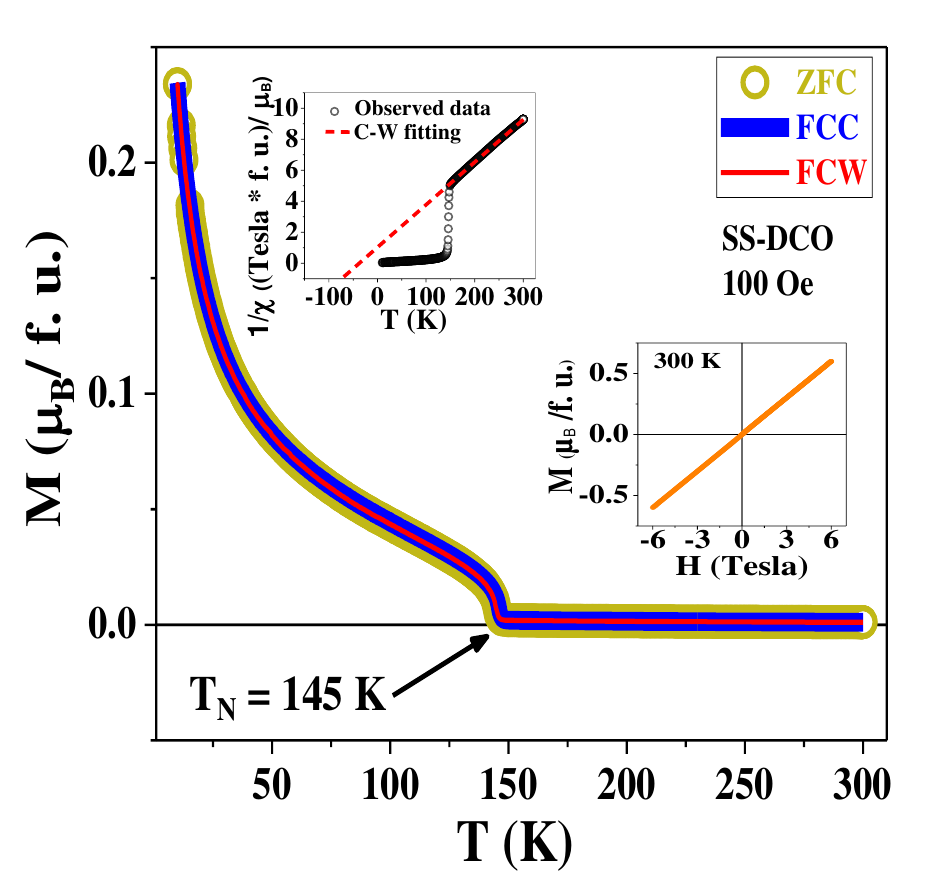}} 
		\vspace*{-0.0 in}\caption{(Color online) Temperature (\textit{T}) dependence of magnetization (\textit{M}) of SS-DCO collected in ZFC, FCC, and FCW modes with applied field (\textit{H}) of 100 Oe, which indicate an antiferromagnetic transition at around 145 K. The upper inset shows the corresponding \textit{1/$\chi$} vs.\textit{ T} data that clearly highlight antiferromagnetic interactions and the lower inset presents linear \textit{M-H} data suggesting the paramagnetic state of SS-DCO at room temperature.}\label{}
	\end{figure}

	\section{X. The room temperature PUND measurements on a separate piece of SPS-DCO after one year}

\renewcommand{\thefigure}{S4}
\begin{figure}[h!]
	\vspace*{-0.15 in}
	\hspace*{-0.0 in}\scalebox{0.8}
	{\includegraphics{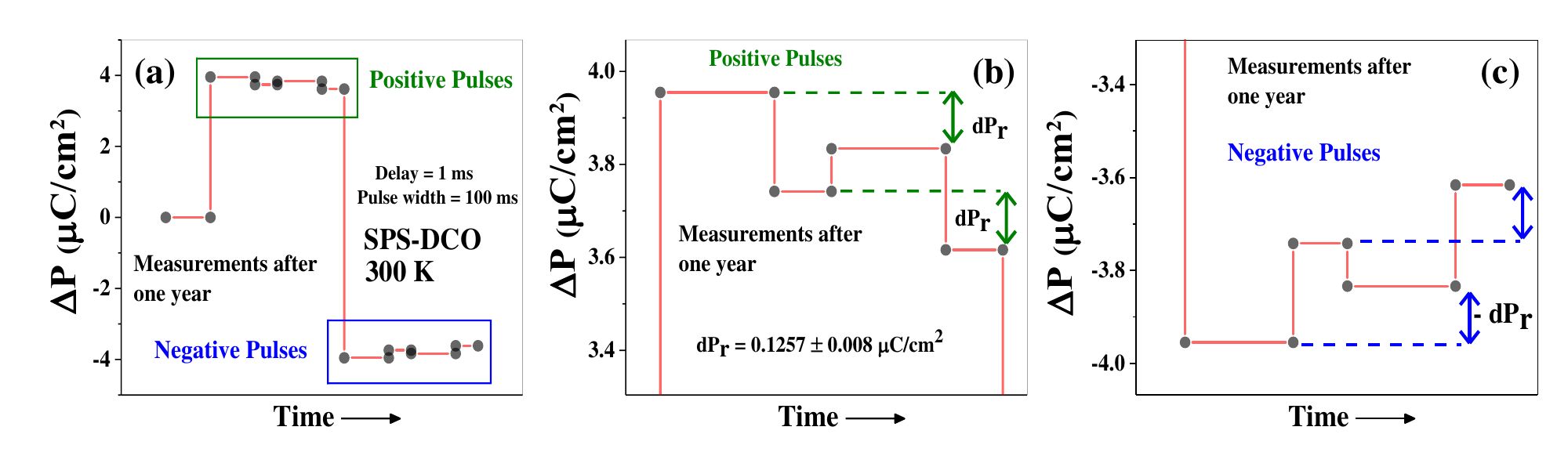}} 
	\vspace*{-0.0 in}\caption{(Color online) (a) Room temperature PUND data at 6 kV/cm applied electric field taken on a separate piece of SPS-DCO after one year. The figures (b) and (c) show enlarged views of the switchable electric polarization responses under positive and negative electric pulses of PUND data.}\label{}
\end{figure}

\newpage
\section{XI. Magnetization (\textit{M}) vs. temperature (\textit{T}) data of SS-LCO}

\renewcommand{\thefigure}{S5}
\begin{figure}[h!]
	\vspace*{-0.15 in}
	\hspace*{-0.0 in}\scalebox{0.4}
	{\includegraphics{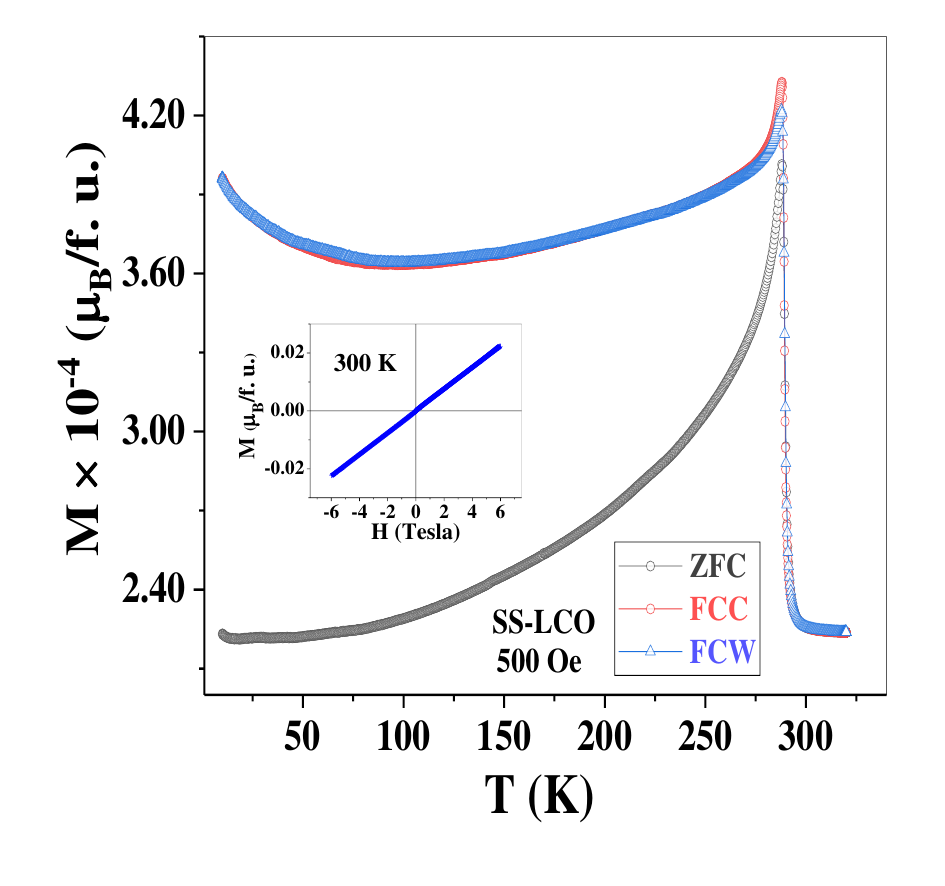}} 
	\vspace*{-0.0 in}\caption{(Color online) ZFC, FCC, and FCW \textit{M}-\textit{T} data of SS-LCO which show a magnetic transition at around 290 K and the inset \textit{M}-\textit{H} data indicates a paramagnetic state at room temperature. }\label{}
\end{figure}

\section{XII. Temperature dependencies of relative dielectric permittivity, dielectric loss, and DSC heating data of SPS-LCO and SS-LCO}

\renewcommand{\thefigure}{S6}
\begin{figure}[h!]
	\vspace*{-0.15 in}
	\hspace*{-0.0 in}\scalebox{0.5}
	{\includegraphics{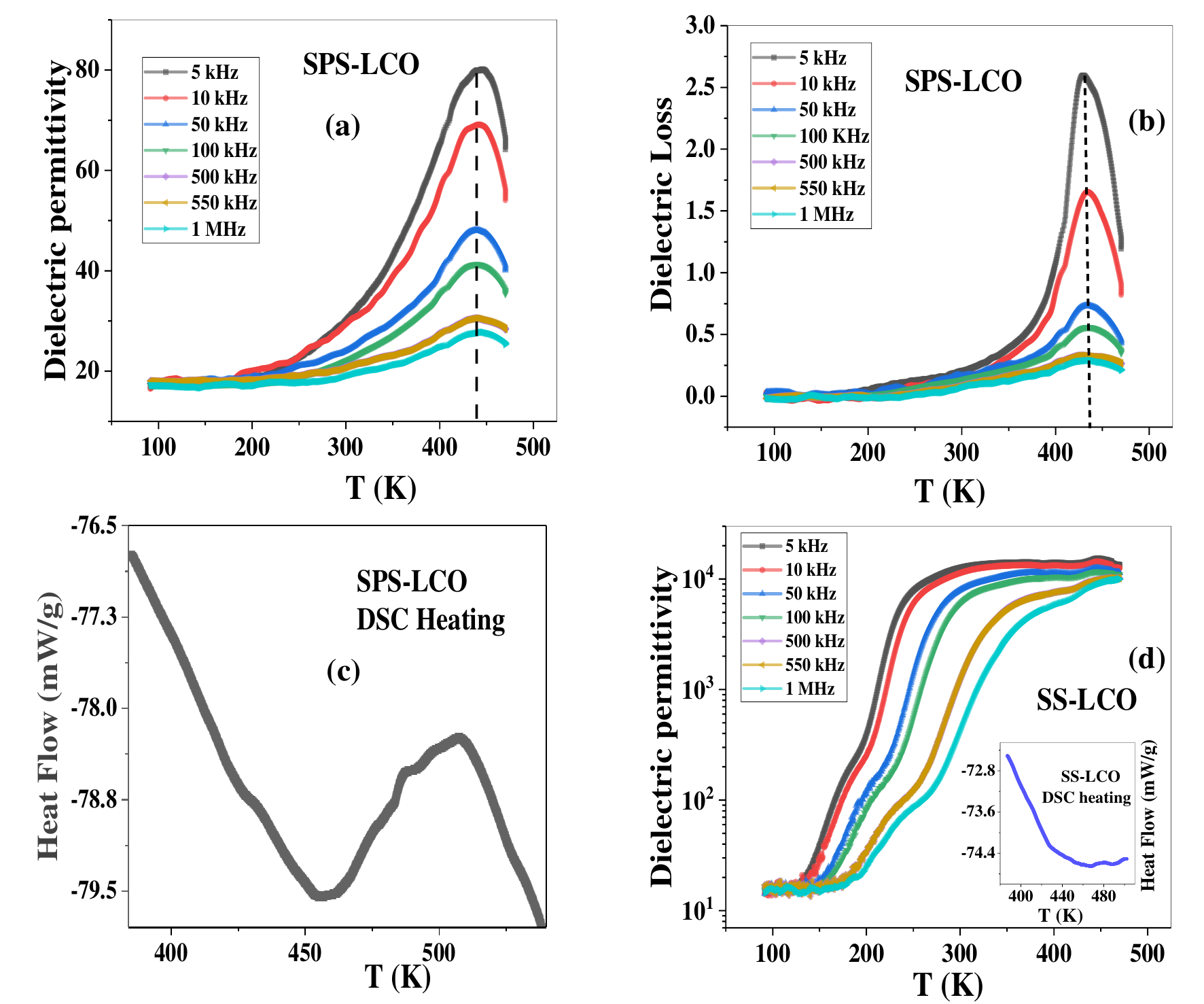}} 
	\vspace*{-0.0 in}\caption{(Color online) (a) Temperature dependencies of (a) relative dielectric permittivity, and (b) dielectric loss data of SPS-LCO which exhibit frequency dispersionless peaks around 440 K. (c) DSC heating data of SPS-LCO also exhibits the similar higher temperature transition. (d) Temperature dependence of relative dielectric permittivity data of SS-LCO, where such a dispersionless peak feature is not found (the inset shows the DSC heating data of SS-LCO}\label{}
\end{figure}

\newpage

\section{XIII. Temperature dependencies of relative dielectric permittivity data collected in 2nd heating cycle, and DSC cooling data of SPS-LCO}

\renewcommand{\thefigure}{S7}
\begin{figure}[h!]
	\vspace*{-0.15 in}
	\hspace*{-0.0 in}\scalebox{0.4}
	{\includegraphics{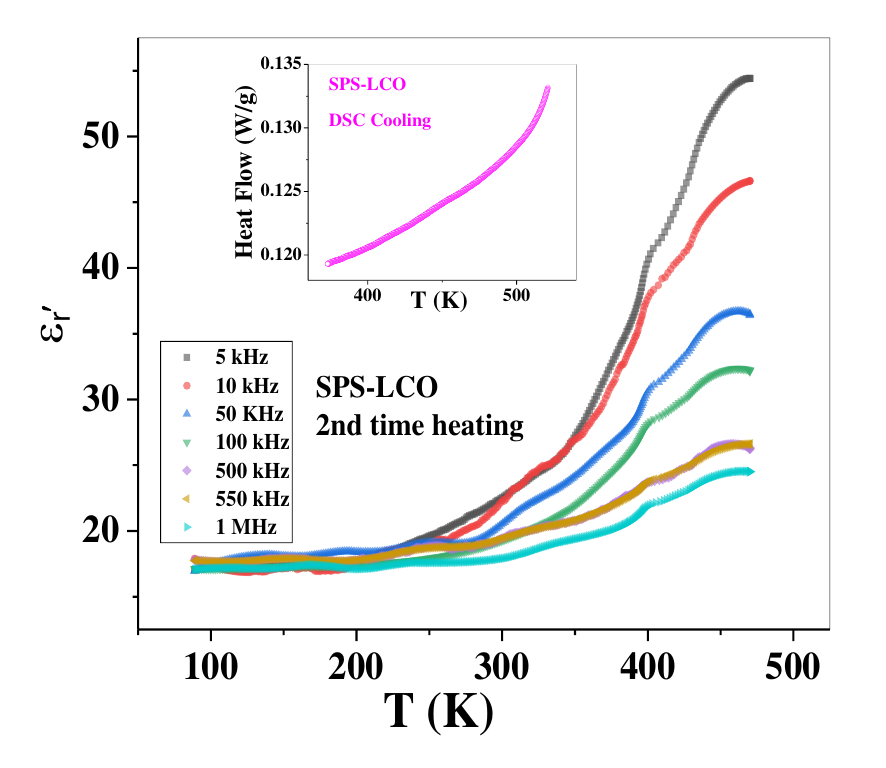}} 
	\vspace*{-0.0 in}\caption{(Color online) Temperature dependence of relative dielectric permittivity data of SPS-LCO taken in the 2nd heating cycle which shows the absence of high temperature transition (the inset shows the DSC cooling data of SPS-LCO).}\label{}
\end{figure}

\section{XIV. Room temperature synchrotron XRD spectra of SPS-LCO and SS-LCO}

\renewcommand{\thefigure}{S8}
\begin{figure}[h!]
	\vspace*{-0.15 in}
	\hspace*{-0.0 in}\scalebox{0.5}
	{\includegraphics{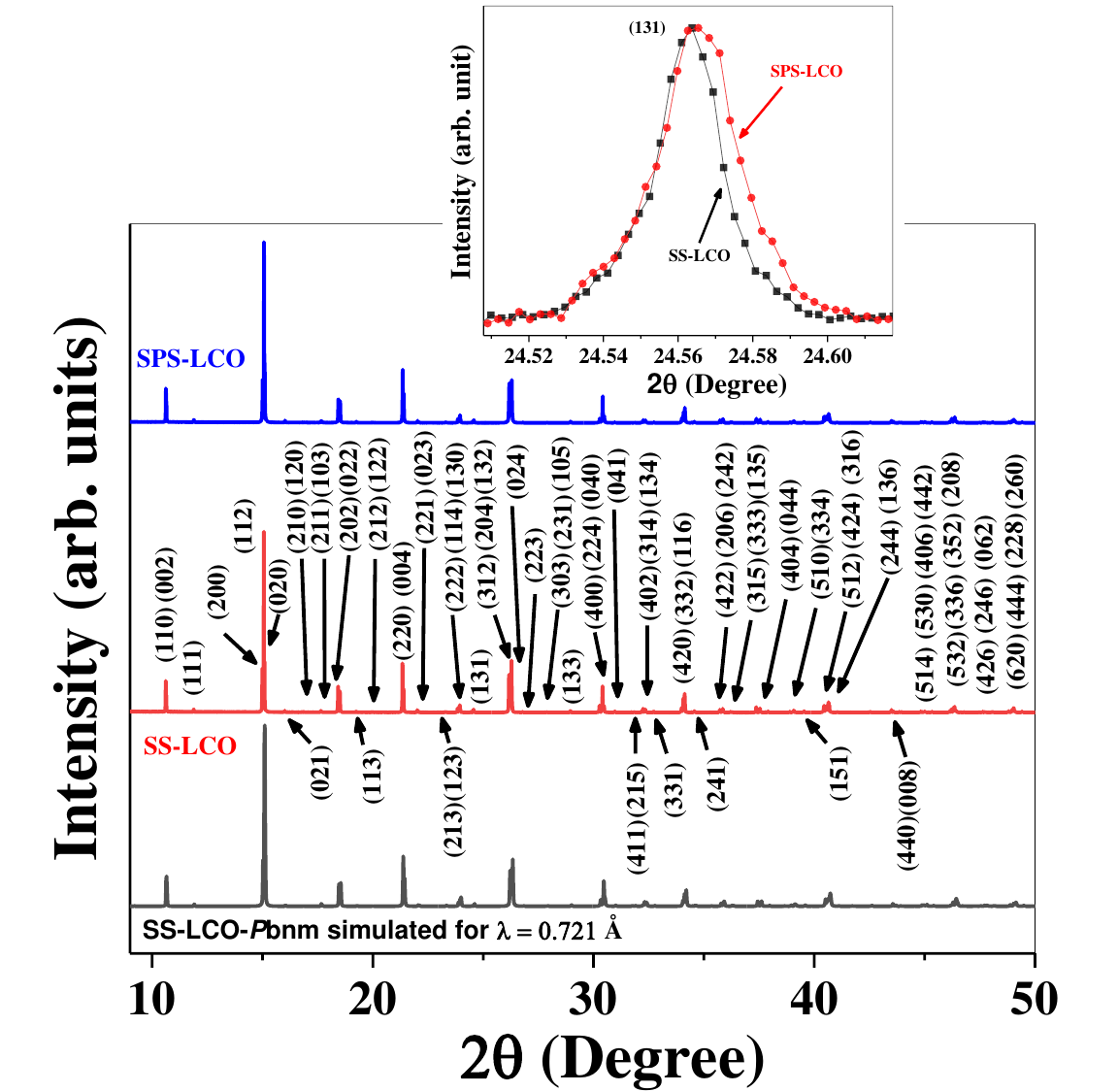}} 
	\vspace*{-0.0 in}\caption{(Color online) Room-temperature synchrotron XRD spectra of SPS-LCO and SS-LCO. The inset shows comparison of one representative peak, which suggest that although structurally similar and single-phase, SPS-LCO does exhibit distinct peak-broadenings as compared to SS-LCO.}\label{}
\end{figure}

\newpage

\section{XV. Rietveld refinement of room temperature synchrotron XRD spectra of SPS-LCO and SS-LCO}

\renewcommand{\thefigure}{S9}
\begin{figure}[h!]
	\vspace*{-0.15 in}
	\hspace*{-0.0 in}\scalebox{0.8}
	{\includegraphics{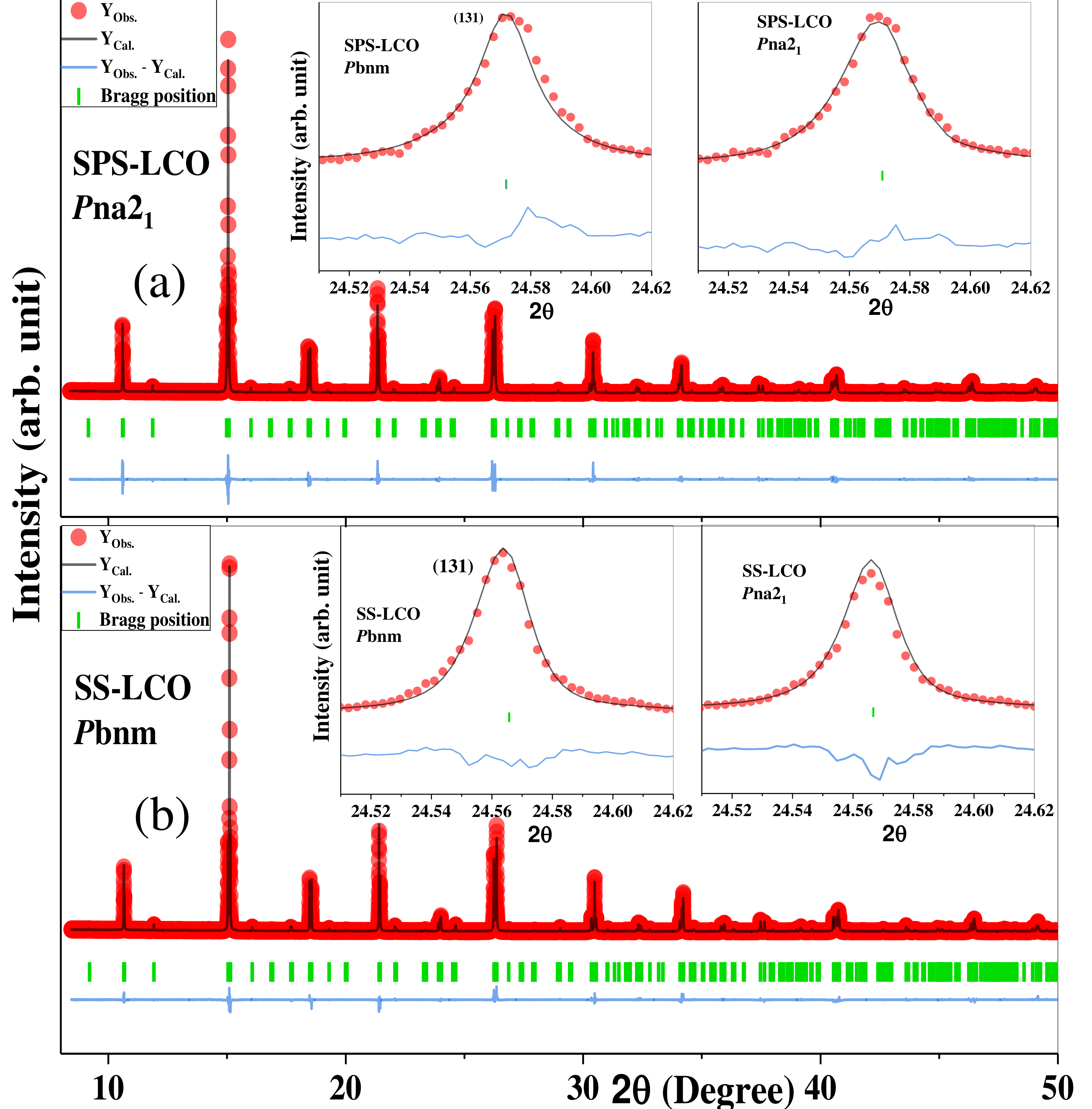}} 
	\vspace*{-0.0 in}\caption{(Color online) (a) The Rietveld refinement of room temperature synchrotron XRD spectra of SPS-LCO considering \textit{P}na2$_1$ space group. The left and right insets shows (131) peak of SPS-LCO, refined with \textit{P}bnm (R$_{\rm{p}}$ = 10.1, R$_{\rm{wp}}$= 13.5, $\chi^{2}$ = 3.81) and with \textit{P}na2$_1$ (R$_{\rm{p}}$ = 9.21, R$_{\rm{wp}}$ = 11.9, $\chi^{2}$ = 2.98) space groups, respectively. (b) The Rietveld refinement of room temperature synchrotron XRD spectra of SS-LCO with \textit{P}bnm space group. The left and right inset shows (131) peak of SS-LCO, refined with \textit{P}bnm (R$_{\rm{p}}$ = 8.53, R$_{\rm{wp}}$ = 10.8, $\chi^{2}$ = 2.76) and with \textit{P}na2$_1$ (R$_{\rm{p}}$ = 9.58, R$_{\rm{wp}}$ = 12.5, $\chi^{2}$ = 3.70) space groups, respectively.}\label{}
\end{figure}

\newpage
 \section{XVI. Lattice parameters obtained from Rietveld refinement of room temperature synchrotron XRD of SPS- LCO and SS-LCO}

\setlength{\tabcolsep}{0.42 cm}
\begin{table}[h]
	\centering
	\caption{}.
	\label{}
	\vspace*{0.5cm}
	\begin{tabular}{|c|c|}
		\hline
	
		SPS-LCO-Pna2$_1$ &  SS-LCO-\textit{P}bnm  \\
		
		\hline
	
		a = 5.476329(20) {\AA}  &  a = 5.508203(11) {\AA}  \\
		
		\hline
		
		b = 5.511812(18) {\AA} & b = 5.472083(12) {\AA}    \\
		
		\hline
	
		c= 7.75341(3) {\AA}  &  c= 7.748075(17) {\AA}\\
		
		\hline
		
	\end{tabular}
	
\end{table}

\section{XVII. Room temperature synchrotron XRD spectra and Rietveld refinement of SPS-LCO-ANN}

\renewcommand{\thefigure}{S10}
\begin{figure}[h!]
	\vspace*{-0.15 in}
	\hspace*{-0.0 in}\scalebox{1.2}
	{\includegraphics{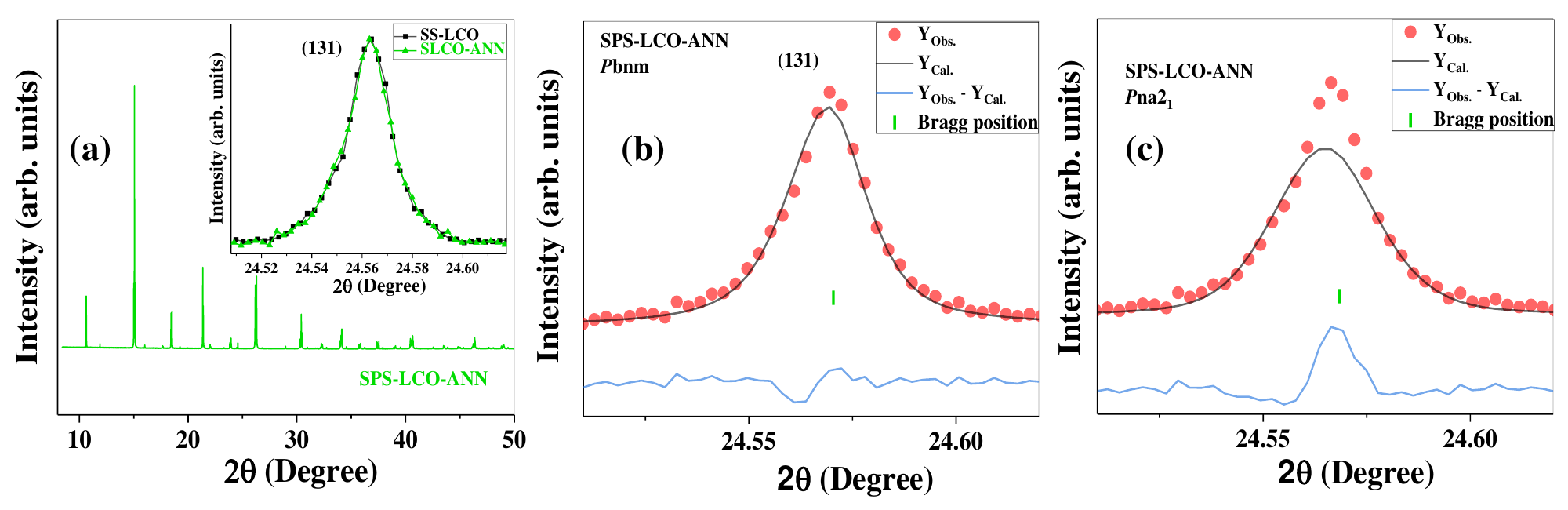}} 
	\vspace*{-0.0 in}\caption{(Color online) (a) Room temperature synchrotron XRD of SPS-LCO-ANN sample. The inset shows the XRD peak comparison between SPS-LCO-ANN and SS-LCO.  The Rietveld refinement of room temperature synchrotron XRD of SPS-LCO annealed sample (SPS-LCO-ANN) with (b) \textit{P}bnm (R$_{\rm{p}}$, R$_{\rm{wp}}$ = 12.2, $\chi^{2}$ = 2.93), (c) \textit{P}na2$_1$ (R$_{\rm{p}}$ = 11.0, R$_{\rm{wp}}$ = 14.2, $\chi^{2}$ = 3.98).}\label{}
\end{figure}

\end{document}